\documentstyle[amssymb,pra,epsfig,aps]{revtex}

\begin{document}
\draft
\title{The Fej\'{e}r Average and the Short Term Behaviors of a Wave Packet in Infinite Square Well}
\author{Quan-Hui Liu$^{1,2,3}$, Wei-Hong Qi$^{1}$, Li-Ping Fu$^{1}$,and Bambi Hu$^{2,4}$}
\address{$^{1}$ Department of Applied Physics, Hunan University, Changsha, 410082, China, \\
$^{2}$ Department of Physics, and the Center for Nonlinear Studies,\\
Hong Kong Baptist University, Hong Kong, China, \\
$^{3}$ Institute of Theoretical Physics, Chinese Academy of Sciences, P.O.
Box 2735, Beijing, 100080, China, \\
$^{4}$ Department of Physics, University of Houston, Houston, TX 77204-5506,
USA}
\date{\today}
\maketitle

\begin{abstract}
The first two period behaviors of a quantum wave packet in an infinite square well potential is studied. First, the short term behavior of expectation value of a quantity on an equally weighted wave packet (EWWP) is in classical limit proved to reproduce the Fej\'{e}r average of the Fourier series decomposition of
the corresponding classical quantity. Second, in order to best mimic the
classical behavior, a nice relation between number $N$ of stationary states
in the EWWP with the average quantum number $n$ as $N\thickapprox \sqrt{n}
$ is revealed. Third, since the Fej\'{e}r average can only approximate the
classical quantity, it carries an uncertainty which in large quantum number
case is almost the same as the quantum uncertainty.
\end{abstract}


{\it Introduction} In quantum mechanics, wave packet is a basic concept
having broad applications \cite{wave}. During the past few years, the
dynamics of the wave packets has become a very active field of research \cite
{wave}. Intensive theoretical and experimental studies show that only the
short term, one or two periods for precise, behaviors of an initially
localized wave pocket prove to be like those of an appropriate ensemble of
classical orbits. Its long term behaviors such as collapse, (super-)revival
etc. are inevitable and of purely quantum mechanical origin \cite{wave}.
However, an outstanding problem remains open: In what sense can or can not
the quantum mechanics recover the classical mechanics for a single orbit in
classical limit?

As stated in standard textbook \cite{lan}, in order to obtain a definite
classical orbit in classical limit, we must start from a semiclassical wave
packet of a particular form $\sum_{m}c_{m}\psi _{m}$, where the coefficients 
$c_{m}$ are noticeably different from zero only in some range $\delta m$ of
values of the quantum number $n$ such that $1\ll \delta m\ll n$; the numbers 
$n$ are supposed large and the superposed states $\psi _{{m}}$ have nearly
the same energy $E_{n}$ \cite{lan}. The mostly utilized wave packet $%
\sum_{m}c_{m}\psi _{m}$ is the Gaussian wave packet in which the
distribution of $m$ in $c_{m}$ is Gaussian. As believed, the expectation
value of a quantity in such a wave packet must become, in the classical
limit, simply the classical value of the quantity; and the expectation value
in the classical limit was ``proved'' to be the Fourier series form of the
classical quantity \cite{lan}. In fact, the proof given in \cite{lan} is not
rigorous in mathematics, as recently shown in \cite{liu}. In fact, if thing
is really so simple, some arguments in quantum-to-classical correspondence
would have been settled long before. There are indeed many delicate problems
awaiting to be resolved.

Even the form is quite simpler than the Gaussian wave packet, the equally
weighted wave packet (EWWP) coveres the essence of the semiclassical wave
packet \cite{liu}. By EWWP, we mean that there are only $2N+1$ stationary
states around the $n$th superposed into the wave packet with the same weight 
$1/\sqrt{2N+1}$. Explicitly the EWWP $|\psi (t)>$ is 
\begin{equation}
|\psi (t)>=\frac{1}{\sqrt{2N+1}}\sum_{m=-N}^{N}\psi _{n+m}(x)\exp (\frac{%
-iE_{n+m}t}{\hbar }),\text{ \ \ \ }(N\prec n).
\end{equation}
We recently proved that in the constraint classical limit $\hbar \rightarrow
0$, $n\hbar =$ an appropriate classical action, the quantum mechanical
average of a quantity on the EWWP goes over to the classical quantity. But
the form is Fej\'{e}r average \cite{gibbs} rather than the Fourier series of
the classical quantity \cite{liu}. In general, for a piecewise smooth
function $f(x)$ on a circle, the Fej\'{e}r average, denoted by $%
F\left\langle f\right\rangle $, converges in the usual sense (i.e., $%
\lim_{x\rightarrow x_{0}}f(x)=f(x_{0})$ ) to, while the truncated Fourier
series of $f(x)$ converges in the mean to, $\left( f(x+0)+f(x-0)\right) /2$ 
\cite{gibbs}. However, the choice of $N$ is not arbitrary: It is determined
by the minimizing the uncertainty relation $\Delta x\Delta p$. The relation
between $N$ and $n$ can be reasonably assumed to be given by 
\begin{equation}
\begin{array}{cc}
N\thicksim n^{m}, & m<1,
\end{array}
\end{equation}
where $m$ is a constant; and the value of $m$ depends on the form of
potential. For instance, the analytical calculation can give for a harmonic
oscillator $m=2/3$, whereas the numerical calculation in this Letter will
show for the infinite square well potential $m=1/2$. The quantum motion of a
single particle in an infinite square well is worthy of paying special
attention, not only because it is a fundamental model in quantum mechanics
and insights into the dynamics exhibited in it speak immediately to a widely
range of physical systems \cite{stroud}, but also the classical position $x$
and momentum $p$ against time $t$ give the sawtooth and square wave
respectively (cf. the Fig. 2), which are indispensable in elementary
discussion of Fourier analyses relevant to Fej\'{e}r average \cite{gibbs}
that offers the base of our exact treatment. This Letter dedicates to the
study of the quantum motion of EWWP in the well.

This Letter starts with a direct illustration that the quantum motion of the
EWWP in an infinite square potential well and its classical limit, where the
Fej\'{e}r averages come out in the following classical limit (\ref{lim}). Next
with the value of the Planck's constant remaining invariant, the behaviors
of the EWWP in first two period are studied. Since EWWP is a semiclassical
wave packet, the requirement on minimizing of the uncertainty relation $%
\Delta x\Delta p$ should be imposed. Numerical result shows that $%
N\thickapprox \sqrt{n}$. This relation enables us to do two things: First,
since the Planck's constant is finite, the quantum number $n$ as well as $N$
is also finite. Direct comparison between quantum mechanical expectation
value, the classical quantity and its Fej\'{e}r average approximation shows
that the Fej\'{e}r average rather than the classical quantity itself is much
closer to the expectation value. Second, we can clearly see that the small
difference between the quantum mechanical average and the Fej\'{e}r average
comes mainly from the inequality of energy level spacing. In this Letter  $%
n=500$ is chosen to demonstrate these two respects. Finally, a brief
conclusion is given.

{\it Basic results} For the quantum motion of a particle of mass $\mu $ in
the one dimension infinite square well with width $a$, the normalized
stationary state function is $\psi _{n}(x)\exp (-iE_{n}t/\hbar )$ $%
=(2/a)^{1/2}\sin (k_{n}x)\exp (-ip_{n}^{2}t/(2\mu \hbar ))$, where $%
k_{n}=n\pi /a$, $E_{n}=p_{n}^{2}/(2\mu )$ is the energy and the momentum $%
p_{n}=\pm \hbar k_{n}$ are equally probable. In classical mechanics, the
particle moves to and fro within the two impenetrable walls. The classical
position $x$ is $p_{c}t/\mu =a\omega t/\pi $, when $0<t<T/2$, and $%
2a-a\omega t/\pi $, when $T/2<t<T$, where $T=2a{\mu }/p_{c}$ is the time of
one period, $p_{c}$ is the magnitude of momentum, and $\omega =2\pi /T$ is
the frequency. The derivative of position $x$ with respect to time $t$ gives
the velocity $p/\mu $. The $m$th partial sum of its Fourier series
representing $x$, or simply called the $m$th truncated Fourier series of $x$ \cite{gibbs}, is given by $x_{m}$ as

\begin{equation}
x_{m}=\frac{a}{2}-\frac{4a}{\pi ^{2}}\sum_{r=0}^{m}\frac{\cos [(2r+1)\omega
t]}{(2r+1)^{2}}.
\end{equation}
Its Fej\'{e}r average $F\left\langle x\right\rangle $ is 
\begin{equation}
F\left\langle x\right\rangle =\frac{a}{2}-\frac{8a}{\pi ^{2}}\frac{1}{(2N+1)}%
\sum_{l=0}^{N-1}\sum_{r=0}^{l}\frac{\cos [(2r+1)\omega t]}{(2r+1)^{2}}.
\end{equation}
Similarly, we have $F\left\langle f\right\rangle $ for $f=x^{2}$, $p,$ $%
p^{2} $ (cf. the following Eqs.(\ref{qx})-(\ref{qps})). To note that there
is the famous Gibb's phenomenon when using truncated Fourier series to
approximate the classical momentum $p$, while $F\left\langle p\right\rangle $
has not such a phenomenon \cite{gibbs}. The calculation of $\ \left\langle
f\right\rangle $ in the EWWP for $f=x,$ $x^{2},$ $p,$ $p^{2}$ is
straightforward, and they are 
\begin{eqnarray}
\left\langle x\right\rangle &=&\frac{a}{2}+\frac{4a}{\pi ^{2}}\frac{1}{2N+1}%
\stackrel{N-1}{%
\mathrel{\mathop{\sum }\limits_{l=0}}%
}\stackrel{l}{%
\mathrel{\mathop{\sum }\limits_{r=0}}%
}\left\{ \left[ \frac{1}{(2n+2N-4l+2r-1)^{2}}-\frac{1}{(2r+1)^{2}}\right]
\cos \left[ (2r+1)\left( 1+\frac{2N-4l+2r-1}{2n}\right) \omega _{n}t\right]
\right.  \nonumber \\
&&+\left. \left[ \frac{1}{(2n+2N-4l+2r-3)^{2}}-\frac{1}{(2r+1)^{2}}\right]
\cos \left[ (2r+1)\left( 1+\frac{2N-4l+2r-3}{2n}\right) \omega _{n}t\right]
\right\} \text{;}  \label{qx}
\end{eqnarray}

\begin{eqnarray}
\left\langle x^{2}\right\rangle &=&\frac{a^{2}}{3}-\frac{1}{2N+1}\frac{a^{2}%
}{2\pi ^{2}}\stackrel{N}{%
\mathrel{\mathop{\sum }\limits_{m=-N}}%
}\frac{1}{(n+m)^{2}}  \nonumber \\
&&+\frac{4a^{2}}{\pi ^{2}}\frac{1}{2N+1}\stackrel{2N}{%
\mathrel{\mathop{\sum }\limits_{l=1}}%
}\stackrel{l}{%
\mathrel{\mathop{\sum }\limits_{r=1}}%
}(-1)^{r}\left[ \frac{1}{r^{2}}-\frac{1}{(2n-2N+2l-r)^{2}}\right] \cos \left[
r\left( 1-\frac{2N-2l+r}{2n}\right) \omega _{n}t\right] \text{;}  \label{qxs}
\end{eqnarray}

\begin{equation}
\left\langle p\right\rangle =\mu \frac{d}{dt}\left\langle x\right\rangle 
\text{;}  \label{qp}
\end{equation}

\begin{equation}
\left\langle p^{2}\right\rangle =\frac{1}{2N+1}\left( \frac{\pi \hbar }{a}%
\right) ^{2}\sum\limits_{m=-N}^{N}(n+m)^{2}=\left( \frac{n\pi \hbar }{a}%
\right) ^{2}(1+\frac{N+N^{2}}{3n^{2}})=p_{n}^{2}(1+\frac{N+N^{2}}{3n^{2}})%
\text{,}  \label{qps}
\end{equation}
where $\omega _{n}=\pi p_{n}/(\mu a)=n\hbar \pi ^{2}/(\mu a^{2})$. Strictly
speaking, only the following set of classical limits 
\begin{equation}
n\rightarrow \infty ,\text{ }N\rightarrow \infty ,\text{ }N/n\rightarrow 0,%
\text{ }n\hbar \rightarrow p_{c}a/\pi =\omega \mu (a/\pi )^{2}\text{ (i.e, }%
\left| p_{n}\right| \rightarrow p_{c}\text{, or, }\omega _{n}\rightarrow
\omega \text{),}  \label{lim}
\end{equation}
is necessary and sufficient to make the quantum mechanical averages (\ref{qx}%
)-(\ref{qps}) exactly equal to the following Fej\'{e}r averages respectively 
\begin{equation}
\left\langle x\right\rangle =F\left\langle x\right\rangle =\frac{a}{2}-\frac{%
8a}{\pi ^{2}}\frac{1}{2N+1}\stackrel{N-1}{%
\mathrel{\mathop{\sum }\limits_{l=0}}%
}\stackrel{l}{%
\mathrel{\mathop{\sum }\limits_{r=0}}%
}\frac{\cos [(2r+1)\omega t]}{(2r+1)^{2}}\text{;}
\end{equation}

\begin{equation}
\left\langle x^{2}\right\rangle =F\left\langle x^{2}\right\rangle =\frac{%
a^{2}}{3}+\frac{4a^{2}}{\pi ^{2}}\frac{1}{2N+1}\stackrel{2N}{%
\mathrel{\mathop{\sum }\limits_{l=1}}%
}\stackrel{l}{%
\mathrel{\mathop{\sum }\limits_{r=1}}%
}\left( -1\right) ^{r}\frac{\cos (r\omega t)}{r^{2}}\text{;}
\end{equation}
\begin{equation}
\left\langle p\right\rangle =F\left\langle p\right\rangle =\mu \frac{d}{dt}%
F\left\langle x\right\rangle \text{;}
\end{equation}

\begin{equation}
\left\langle p^{2}\right\rangle =\left( \frac{n\pi \hbar }{a}\right) ^{2}(1+%
\frac{N+N^{2}}{3n^{2}})\stackrel{in\text{ }(\ref{lim})}{=}p_{n}^{2}\text{.}
\end{equation}
In fact, we have recently proved in general that the classical limit of mean
value of every quantity on an EWWP can exactly give the Fej\'{e}r average of
the Fourier series expansion of its corresponding classical quantity \cite
{liu}. So far, we can draw with safety the following conclusion: In
conformity with mathematical rigor, quantum mechanics can never reproduce
the exact classical mechanics unless in the nonphysical limit $\hbar
\rightarrow 0$. This situation is not new to us. The special relativity can
also not reduce to the classical Newtonian mechanics unless in the
nonphysical limit the speed of light $c\rightarrow \infty $.

{\it Large quantum number limit only is not sufficient to reproduce the
classical mechanics from quantum mechanics} In the previous section, we have
demonstrated that the single classical orbit can not be recovered unless in
the limit (\ref{lim}) involving the nonphysical limit $\hbar \rightarrow 0$.
In physical sense, this limit $\hbar \rightarrow 0$ can only be viewed as
the comparably small to the classical action, which is essentially the large
quantum number case. But the large quantum number alone is not sufficient to
reproduce the classical mechanics from quantum mechanics. Its long term
behaviors such as collapse, (super- or fractional-) revival etc. are doomed
to appear \cite{stroud}. Even in the short term, a finite $n$ means a finite 
$N$ as well; and a EWWP with finite $2N+1$ stationary states approaches to
the Fej\'{e}r average of $\ $the first $2N$ $+1$ partial sums of the Fourier
series decomposition of the classical quantity only. The finite term
Fej\'{e}r average can in essence approximate the classical quantity.
Therefore the large quantum number limit is in general not sufficient to
reproduce the classical mechanics from quantum mechanics. In this section,
we will study these problems in detail. In all numerical calculations, the
natural units are used in which $a=\mu =\hbar =1$.

To give the relation between $N$ and $n$ is easy, which can be obtained by
minimizing the uncertainty relation $\Delta x\Delta p$. To note that $%
\left\langle p\right\rangle $ is a continuous function of $\ t$; it is zero
when the center of EWWP starts to return from either of the two walls. Since
the expectation of $p^{2}$ is a constant of time, $\Delta p$=$\sqrt{%
\left\langle p^{2}\right\rangle }$ when the $\left\langle p\right\rangle =0$%
. At this instant, minimizing $\Delta x\Delta p$ amounts to minimizing $%
\Delta x$. Numerical result as given in Fig. 1 shows that $N\thickapprox 
\sqrt{n}$, e.g., $N=23$ when $n=500$. In Fig.2-Fig.4, various quantities in
first two periods are plotted when $n=500$ $\ (N=23),\ p_{c}=500\pi $, $%
T=0.00127$.

As shown in Fig. 2, there is little difference between $F\left\langle
f\right\rangle $ and $\left\langle f\right\rangle $ for $f=x,$ $p$
respectively. In order to yield a marked difference between the quantum
mechanical expectation value, Fej\'{e}r average and the represented
function, a reduced uncertainty $\delta f$ \ defined as $\delta f\equiv $ $%
\sqrt{1-\left\langle f\right\rangle ^{2}/\left\langle f^{2}\right\rangle }$
and the classical one $\ \delta ^{\prime }f$ \ defined as $\ \delta ^{\prime
}f\equiv $ $\sqrt{1-(F\left\langle f\right\rangle )^{2}/F\left\langle
f^{2}\right\rangle }$ \ are introduced. In Fig.3 and Fig.4, we compare the
behaviors of $\delta ^{\prime }f$ and $\delta f$ for $f=x$, $p$
respectively. We use solid lines to plot the reduced classical uncertainties, and the dotted for either the pure quantum mechanical quantities or the
quasi-quantum mechanical ones. By the quasi-quantum mechanical quantities,
we mean those constructed from the pure quantum mechanical ones in which the
quantum mechanical matrix elements $f_{n+r,n}=\int_{0}^{a}\psi
_{n+r}(x)f\psi _{n}(x)dx$ are replaced with the $r$th Fourier amplitudes $%
f_{r}$ of the classical quantity $f$, while the time factors $\exp \left\{ 
(E_{n+r}-E_{n})t/\hbar \right\} $ remain unchanged. From Eqs. (\ref{qx})-(%
\ref{qp}), the pure quantum mechanical quantities and the quasi-ones have
little difference when $n$ is large and $t$ is small, e.g. $n=500$, $t\leq
0.0025$. Therefore when we speak of one of them, both are practically
referred to. Thus, the difference between solid and dotted lines comes from
the inequality of energy level spacing. The longer the time evolves, the
bigger the quantum mechanics deviates from the classical mechanics.

In classical mechanics, there is no uncertainty; and the reduced uncertainty
is zero except when $x=0$. In Figs.3-4, for $f=x$, $p$ respectively, $\delta
f$ and $\delta ^{\prime }f$ are almost the same and are not zero at all. It
means that the wave nature can be described by Fej\'{e}r average; namely, Fej%
\'{e}r average accompanies an uncertainty. But this uncertainty is entirely
different from the standard deviation in ensemble statistics. In current
statistical interpretation of quantum mechanics, the uncertainty is usually
interpreted to have a statistical origin \cite{ball}. This assertion
holds true because a wave packet usually corresponds to an ensemble of
classical orbit. In our approach, the uncertainty exists because the
incomplete description of Fej\'{e}r average approximation of the classical
quantity. This uncertainty is unrevealed before because we have seldom tried
to carefully examine the relation between wave packet and a single classical
orbit.

{\it Conclusion} The Fej\'{e}r average was firstly an intellectual creation
in pure mathematics is successfully demonstrated to exist in physics that
reflects the objective structure of nature. The quantum mechanical average
of a quantity on the EWWP in the classical limit goes over to the Fej\'{e}r
average of Fourier series expansion of the classical quantity, rather than
the Fourier series itself as widely accepted. In the limit of large quantum
number, the inequality of the energy level spacing accounts for the small
difference between the quantum mechanical expectation value of a quantum
quantity and the Fej\'{e}r average of the classical quantity. The short term
behaviors of reduced quantum motion uncertainty are almost identical to the
reduced classical ones defined via the Fej\'{e}r average approximation of
classical quantity. This uncertainty is entirely different from either the
classical one (zero) or the statistical standard deviation in the statistical
interpretation of quantum mechanics.

{\it Acknowledgment} This subject is supported in part by National Natural
Science Foundation of China through grant No: 19974010; in part by grants
from the Hong Kong Research Grants Council (RGC), and the Hong Kong Baptist
University Faculty Research Grant (FRG).

\newpage

\vspace*{0.8cm} 
\begin{figure}
\centerline{\psfig{file=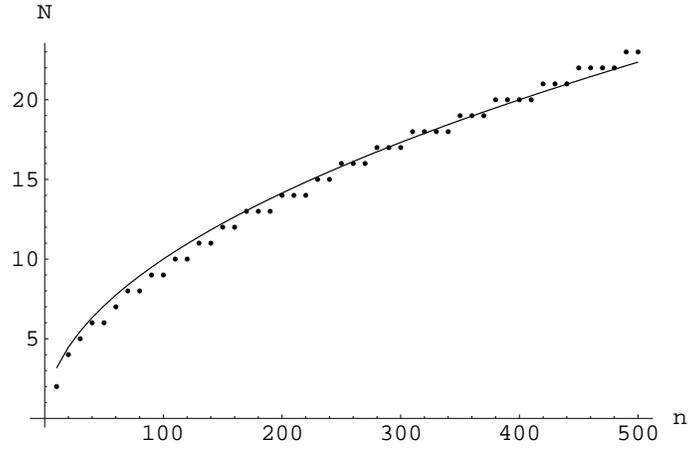,height=6cm}}
\vspace*{1.0cm} 
\caption{Corresponding to each $n$ ranging from $10$ to $500$, there is a unique value of $N$ (dotted curve) minimizing the uncertainty $\Delta x \Delta p$. The smooth curve is $\sqrt{n}$. Since $N=[\sqrt{n}]\pm 1$ with $[x]$ the integer part of $x$, $N\approx\sqrt{n}$ give a nice fit.}
\label{fig1}
\end{figure}

\vspace*{0.8cm} 
\begin{figure}
\centerline{\psfig{file=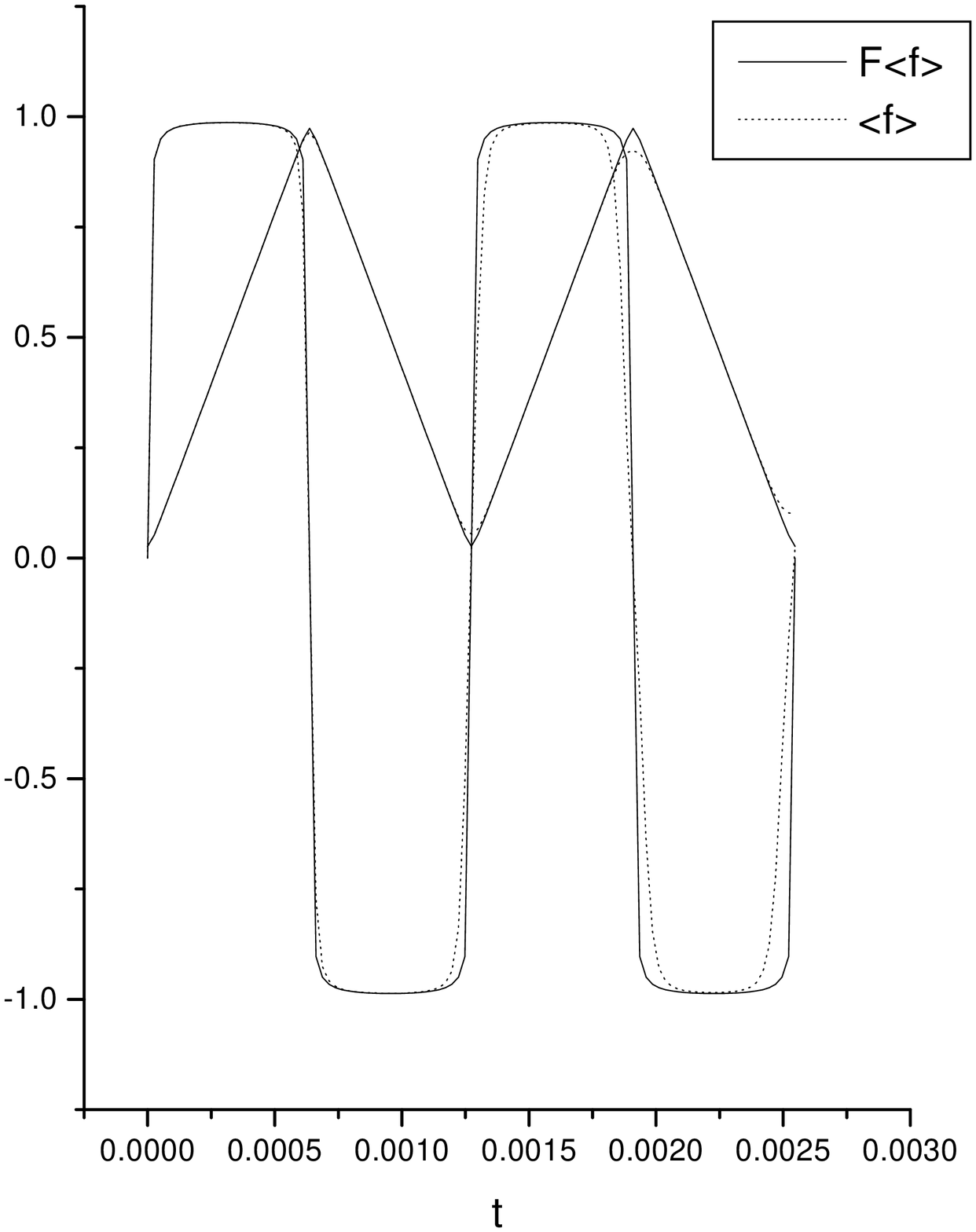,height=7.18cm,width=10cm}}
\vspace*{1.0cm} 
\caption{When $N=23$, $p_{n}=\pm p_{c}=\pm 500\pi $, i.e., $n=500$, and taking $500 \pi$ as the unit of the momentum, the Fej\'{e}r average approximations $F\left\langle x\right\rangle $ (solid sawtooth wave), $F\left\langle p\right\rangle $ (solid square wave). and the quantum mechanical averages $\left\langle x\right\rangle $ (dotted sawtooth curve), and $\left\langle p\right\rangle $ (dotted square curve) in the EWWP. In the first period, $\left\langle x\right\rangle $ and $F\left\langle x\right\rangle $ are almost the same, and so are $\left\langle p\right\rangle $ and $F\left\langle p\right\rangle $. In some intervals two curves completely coincide, only one curve is possibly visible.}
\label{fig2}
\end{figure}

\vspace*{0.8cm} 
\begin{figure}
\centerline{\psfig{file=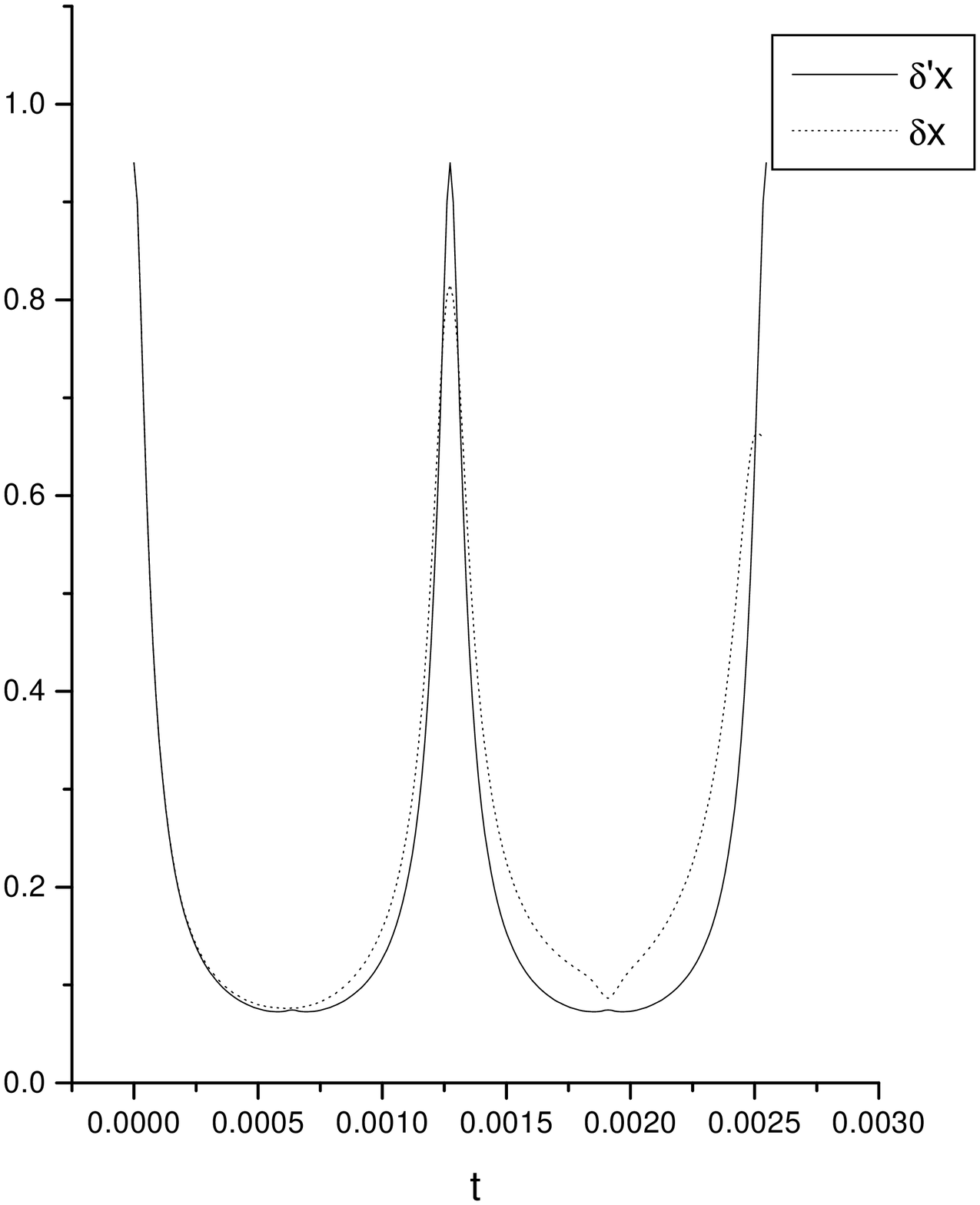,height=7.18cm,width=10cm}}
\vspace*{1.0cm} 
\caption{The difference between $\left\langle x\right\rangle $ and $F\left\langle x\right\rangle $ viewed from the ``magnifier" the reduced uncertainty $\delta x$ (dotted curve) and the classical one $\delta 'x$ (solid curve). The difference between $\left\langle x\right\rangle $ and $F\left\langle x\right\rangle $ comes from the 
inequality of energy level spacing. Both curves are entirely different from classical result $\delta x=0$ ($x\not=0$).}
\label{fig3}
\end{figure}

\vspace*{0.8cm} 
\begin{figure}
\centerline{\psfig{file=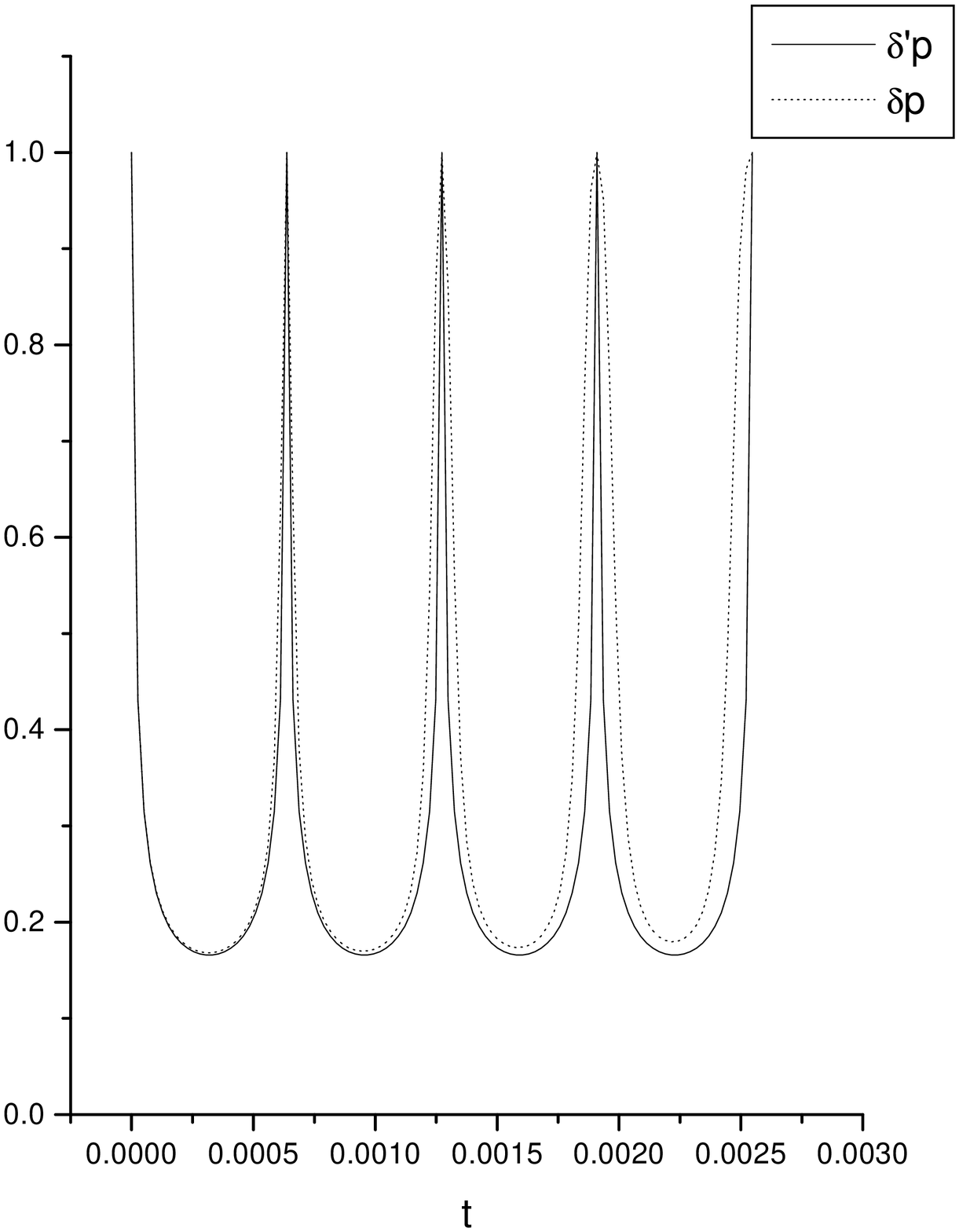,height=7.18cm,width=10cm}}
\vspace*{1.0cm} 
\caption{The difference between $\left\langle p\right\rangle$ and $F\left\langle p\right\rangle $  viewed from the ``magnifier" the reduced uncertainty $\delta p$ (dotted curve) and the classical one $\delta 'p$ (solid curve). Both are entirely different from classical result $\delta p=0$ ($p\not=0$).}
\label{fig4}
\end{figure}

\end{document}